\documentclass[a4paper]{article}

\pdfoutput=1

\usepackage{INTERSPEECH2022}

\usepackage{xcolor}
\usepackage{multirow}
\usepackage[normalem]{ulem}

\usepackage{graphicx}
\usepackage{subfig}
\usepackage{array}
\usepackage{balance}

\newcolumntype{M}[1]{>{\centering\arraybackslash}m{#1}}

\title{Creating New Voices using Normalizing Flows}

\name{Piotr Biliński, Thomas Merritt, Abdelhamid Ezzerg, Kamil Pokora, Sebastian Cygert, Kayoko~Yanagisawa, Roberto~Barra-Chicote, Daniel Korzekwa}
\address{Amazon Alexa}
\email{\{bilipiot, thommer, ezzerg, kamipoko, scygert, yakayoko, rchicote, korzekwa\}@amazon.com}

\begin{document}

\maketitle
\begin{abstract}%
Creating realistic and natural-sounding synthetic speech remains a big challenge for voice identities unseen during training. As there is growing interest in synthesizing voices of new speakers, here we investigate the ability of normalizing flows in text-to-speech (TTS) and voice conversion (VC) modes to extrapolate from speakers observed during training to create unseen speaker identities. Firstly, we create an approach for TTS and VC, and then we comprehensively evaluate our methods and baselines in terms of intelligibility, naturalness, speaker similarity, and ability to create new voices. We use both objective and subjective metrics to benchmark our techniques on 2~evaluation tasks: zero-shot and new voice speech synthesis. The goal of the former task is to measure the precision of the conversion to an unseen voice. The goal of the latter is to measure the ability to create new voices. Extensive evaluations demonstrate that the proposed approach systematically allows to obtain state-of-the-art performance in zero-shot speech synthesis and creates various new voices, unobserved in the training set. We consider this work to be the first attempt to synthesize new voices based on mel-spectrograms and normalizing flows, along with a comprehensive analysis and comparison of the TTS and VC modes.
\end{abstract}

\section{Introduction}
Recent advances in speech synthesis has shown that current technology can produce realistic and natural-sounding synthetic speech for voice identities observed during training~\cite{shen-arxiv-2020,ren2021fastspeech,ezzerg2021enhancing,EDP}. However, speech synthesis with arbitrary new speakers, unseen during training and without post-training adaptation, remains a big challenge \cite{autovc,SCGlowTTS,speakergeneration}. Thus, when we need to generate a new voice, this often means collecting a large high-quality database of a target speaker to train a generative model. This is time-consuming, expensive to record and challenging to enroll speakers even for data reduction speech synthesis methods~\cite{EDP, huybrechts2021low}. Moreover, creating such a database becomes even more difficult if the technology requires a parallel corpus with the same text being read by source and target speakers. Therefore, in this work we tackle the challenging problem of creating new voices using non-parallel data. Our goal is to extrapolate from speakers observed during training to create unseen speaker identities without any recordings of those speakers, and therefore without any target speaker adaptation.

Existing speech synthesis techniques are usually either text-to-speech (TTS) \cite{ren2021fastspeech,Tacotron2,ren2019fastspeech} or voice conversion (VC) \cite{nonparallel-2006,Blow,ours} methods.  Multi-speaker TTS \cite{huybrechts2021low,li2021light} is the process of generating lifelike speech of a target speaker given input text. VC is the process of taking speech from a source speaker and making it sound like the target speaker without changing the linguistic information. While most papers are about either TTS or VC, here we investigate both types of techniques together to discover which is better for generating new voices.

Our investigations are mainly based on normalizing flows, due to their state-of-the-art quality in TTS and increasing interest in the community. These generative models learn the mapping of the input data to a latent space. This mapping is done through a sequence of invertible transformations using the rule of change of variables to obtain a valid probability distribution, allowing for exact sampling and density evaluation~\cite{NF_Intro_Review, GlowTTS, FlowPP}.

Whilst there have been many previous works investigating the application of normalizing flows for speech, these have typically focused on speech synthesis with speakers observed in the training set~\cite{Blow, ours, GlowTTS, FlowTTS, Flowtron}. Moreover, they have typically focused on TTS~\cite{SCGlowTTS, GlowTTS, FlowTTS, Flowtron} or neural vocoding~\cite{ParallelWaveNet, WaveGlow} rather than VC~\cite{Blow, ours}.

Deviating from such approaches, in this work we investigate the ability of normalizing flows in TTS and VC modes to extrapolate from speakers observed during training to create unseen speaker identities. We study the architecture based on Flow-TTS, which has shown to achieve state-of-the-art quality in TTS~\cite{FlowTTS}.
We define Flow-VC as an extension of the TTS model to the voice conversion task.
Then, we comprehensively evaluate our methods and baselines in terms of intelligibility, naturalness, speaker similarity, and ability to create new voices.
We use both objective and subjective metrics to benchmark our techniques on 2 evaluation tasks: zero-shot speech synthesis and new voice speech synthesis. The goal of the former task is to measure the precision of the conversion to an unseen voice. The goal of the latter is to measure the ability of creating new voices.
Extensive evaluations demonstrate that the Flow-VC systematically allows to obtain state-of-the-art performance in zero-shot speech synthesis and creates new voices, unobserved in the training set.
We consider this work to be the first attempt to synthesize new voices based on mel-spectrograms and normalizing flows, along with comprehensive formal evaluations, analysis and comparison of the TTS and VC modes together, and with the state-of-the-art.

\section{Approach}\label{sec:approach}

\begin{figure*}
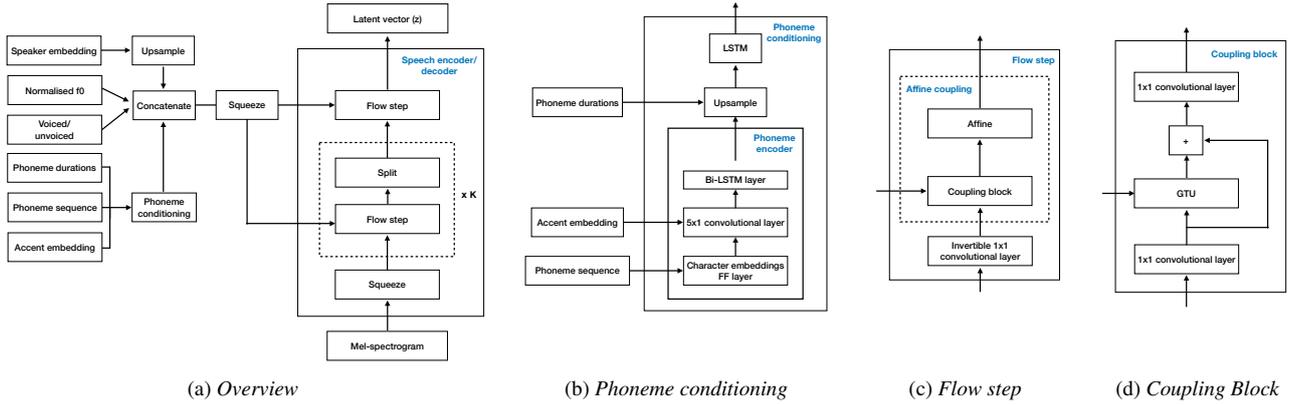
%
\centering
\subfloat[\centering Overview]{{\includegraphics[clip, trim=0cm 0cm 68.6cm 0cm,height=140px]{imgs/1.pdf}}}%
~ ~ ~ ~
\subfloat[\centering Phoneme conditioning]{{\includegraphics[clip, trim=44.4cm 0cm 39.5cm 0cm,height=140px]{imgs/1.pdf}}}%
~ ~ ~ ~
\subfloat[\centering Flow step]{{\includegraphics[clip, trim=73.5cm 0cm 19.5cm 0cm,height=140px]{imgs/1.pdf}}}%
~ ~ ~ ~
\subfloat[\centering Coupling Block]{{\includegraphics[clip, trim=93.5cm 0cm 0cm 0cm,height=140px]{imgs/1.pdf}}}%
\caption{Overview of our TTS and VC approach (a) with phoneme conditioning (b), Flow step (c) and Coupling block (d) components.}%
\label{fig:overview}%
\vspace{-.12cm}
\end{figure*}

Normalizing Flows are generative models with tractable distributions, where both sampling and density evaluation can be both exact and efficient. Our approach is based on \mbox{Flow-TTS}~\cite{FlowTTS}, a non-autoregressive TTS model using normalizing flows, which has shown to achieve state-of-the-art quality in TTS, and thus was selected as the basis for the model topology of the proposed approach.
The main differences between our approach and Flow-TTS are: (i) we tackle the problem of creating new voices rather than TTS synthesis with seen voices, (ii)~we introduce VC mode, (iii) we have the ability to sample new speakers, (iv) we use pre-trained phone alignments, and (v) we use different conditioning to the model. Moreover, we present comprehensive evaluations and analysis of TTS and VC models, and their comparison to the state-of-the-art.

\noindent \textbf{Architecture.} The architecture of our model is presented in Figure~\ref{fig:overview}.
Our model consists of a sequence $f$ of invertible transformations $f_1$, $f_2$, \dots, $f_k$ mapping input mel-spectrogram $m \in \mathbb{R}^{t \times d}$ with $t$ frames and $d$ frequency bins to a latent representation $z \in \mathbb{R}^{t \times d}$ given conditions $\theta$.
There are 6 conditions $\theta = (s, f0, vuv, ph_{seq}, ph_{dur}, a)$:
(i) $s$ is the speaker representation extracted using~\cite{GE2E},
(ii) $f0$ is the sentence-level mean normalised interpolated log-f0 to disentangle speaker identity (speaker's average f0) from sentence prosody and thus separating $f0$ from speaker embedding,
(iii) $vuv$ is a binary 'voiced or unvoiced' flag denoting whether a frame is voiced or unvoiced,
(iv) $ph_{seq}$ is the phoneme sequence representation,
(v) $ph_{dur}$ is the phoneme duration representation,
and (vi) $a$ is the accent embedding, extracted using the approach in~\cite{GE2E}, to allow the model to disambiguate phonemes according to the accent being used.
Conditioning features are passed to the flow steps to help the model better maximise the likelihood. The motivation behind this is that the network can learn to remove the contributions of these conditioning features from the
input data, hence removing the level of noise to the prior. The invertible nature of the flow ensures that the network is information-preserving, meaning that the latent vector is a lossless representation of the input data.
In Flow-TTS, the phoneme alignment is learnt during training using attention, and in Glow-TTS~\cite{GlowTTS}, this is done using monotonic alignment search (MAS). Unlike in these papers, we use pre-trained phone alignments to label the data to simplify the training process, as in attention-free TTS~\cite{EDP, ours}.

\noindent \textbf{Training.} Normalizing Flows model the true data distribution providing an exact likelihood estimate.
The model uses the provided conditions $\theta$ to better map given mel-spectrograms $m$ to a simple known distribution $p_z$ by maximizing the likelihood:
\begin{equation}
\max_f \frac{1}{|M|} \sum_{m \in M} p_m(m),
\end{equation}%
\begin{equation}
p_m(m) = \log p_z(f(m, \theta)) + \log \left| \det \frac{df(m, \theta)}{dm} \right|,
\end{equation}
where $M$ is the set of training mel-spectrograms and $p_z(z)$ is the given prior density, e.g. standard normal distribution.

\noindent \textbf{Flow-TTS.} The model consists of invertible transformation $f$, which allows for lossless reconstruction of mel-spectrogram $m$ from a latent space $z$ given conditions $\theta$.
In text-to-speech (Flow-TTS), we sample $z$ from the prior distribution and run inverse transformation $f$, denoted as $f^{-1}$, to synthesize mel-spectrogram $m_{TTS}$ as $f^{-1}(z, \theta)$.

\noindent \textbf{Flow-VC.} Here, we define the task of unseen voice synthesis as voice conversion task.
We map mel-spectrogram $m$ to a latent representation $z$ given conditions $\theta_{s=s_A}$ (using $s_A$ speaker representation), and then we convert the obtained latent representation $z$ to mel-spectrogram $m_{VC}$ given $\theta_{s=s_B}$ (using $s_B$ speaker representation): $m_{VC} = f^{-1}( f(m, \theta_{s = s_A}), \theta_{s = s_B})$.

\noindent \textbf{New voices generation.}
To generate new voices, we need to provide new speaker embeddings.
These embeddings can be defined arbitrarily, like in zero-shot speech synthesis, or we can automatically generate new representations.
Our approach for generation of new speakers is based on~\cite{speakergeneration}.
The neural network takes as an input a locale embedding and outputs parameters of a Gaussian Mixture Model (GMM) for each dimension of the speaker embedding. We use 10 components for GMM. To generate new voices, for each dimension of the speaker embedding we independently sample GMM using predicted parameters. Our architecture is a simple feed-forward neural network with 2 hidden layers, each of size 256.

\section{Experiments}
We comprehensively evaluate our methods and baselines in terms of intelligibility, naturalness, speaker similarity, and ability to create new voices.
We benchmark our techniques on 2~evaluation tasks: zero-shot and new voice speech synthesis.

\subsection{Evaluation systems}
We evaluate Flow-TTS and Flow-VC models.
The \mbox{Flow-TTS} does not use oracle f0 and vuv conditioning, whereas the \mbox{Flow-VC} does.
We also study the Flow-TTS with additional f0 and vuv conditioning, called Flow-TTS with f0, as well as Flow-VC without f0 and vuv conditioning, called Flow-VC w/o f0.
Moreover, we compare our models with the state-of-the-art External Duration Non-Autoregressive TTS (\mbox{EDNA-TTS})~\cite{EDP} and Tacotron 2~\cite{Tacotron2} TTS models and CopyCat~\cite{CopyCat} VC model.
We use Universal Neural Vocoder~\cite{UniversalNeuralVocoder} based on Parallel WaveNet to map generated mel-spectrograms to audio samples.

\subsection{Dataset}
Our training dataset consists of English recordings of 6 regional accents (American, Australian, British, Canadian, Indian and Welsh). There are 3173 speakers and approximately 579 hours of recordings. The data features a range of different recording conditions with some speakers recorded in studio-quality conditions whilst other speakers were recorded in more ambient surroundings using lower quality microphones.
A sampling rate of 16 kHz was used for all recordings to match the lowest sampling rate observed in the training data, from which 80 dimensional mel-spectrograms were extracted using a frame length of 50 ms and a frame shift of 12.5 ms.

For zero-shot speech synthesis, 6 target speakers were selected (2 American, 2 Australian and 2 Canadian), where 4~speakers are from the VCTK~\cite{VCTK_OLD, VCTK} and 2 speakers are from an internal dataset. These speakers do not exist in training and validation sets.
For VC, we select 60 random utterances for each target speaker, thus we have 360 utterances in total.
For new voice speech synthesis, we use the same 120 American source utterances as for the zero-shot speech synthesis.

\subsection{Zero-shot: Intelligibility}
\begin{table}[t!]
    \centering
    \caption{Zero-shot: WER with 95\% confidence intervals, as well as mean and standard deviation of SECS.}
    \vspace{-.25cm}
    \begin{tabular}{cM{2.2cm}M{2.2cm}}
    \toprule
    \textbf{System} & \textbf{WER}$\downarrow$ & \textbf{SECS} $\uparrow$ \\ \midrule
        Source recordings & ~ 8.99 $\pm$ 1.49 & 0.20 $\pm$ 0.11 \\
        CopyCat & 10.63 $\pm$ 1.61 & 0.51 $\pm$ 0.10 \\
        Tacotron 2 & 17.96 $\pm$ 2.00 & 0.41 $\pm$ 0.12 \\
        EDNA-TTS & ~ 9.69 $\pm$ 1.54 & 0.37 $\pm$ 0.09 \\
        Flow-TTS & 11.37 $\pm$ 1.66 & 0.57 $\pm$ 0.07 \\
        Flow-TTS with f0 & 10.42 $\pm$ 1.59 & 0.53 $\pm$ 0.07 \\
        Flow-VC w/o f0 & 10.01 $\pm$ 1.57 & 0.53 $\pm$ 0.09 \\
        Flow-VC & ~ 9.54 $\pm$ 1.53 & 0.51 $\pm$ 0.09 \\
        \bottomrule
    \end{tabular}
    \label{tab:WER}
    \vspace{-.12cm}
\end{table}%

In order to measure the intelligibility of generated speech, we measure the average Word Error Rate (WER) between the sentence text and the transcription, which we obtain using the AWS Transcribe ASR system.
A US-English ASR model is used for American and Canadian speakers, whilst an Australian-English ASR model is used for Australian speakers.
The WER results for zero-shot evaluation are shown in Table~\ref{tab:WER}. 
The stand-out from these results are that the intelligibility of Tacotron 2 is significantly worse than all other systems. This is likely due to the attention mechanism struggling with the diverse nature of the dataset used. 
Meanwhile the flow models, EDNA-TTS and CopyCat all achieve similar scores, with the lowest WER obtained by Flow-VC.

\subsection{Zero-shot: Flow-TTS vs. Flow-VC}\label{sec_FlowTTS_FlowVC}
The goal of the zero-shot speech synthesis evaluation is to measure the precision of creating unseen voices.
In this experiment, we evaluate and compare Flow-TTS with Flow-VC, both with and without f0 conditioning to understand the impact of: 1)~keeping the latent vector from the source speaker vs. sampling a new latent vector, and 2) using f0 conditioning.

We evaluate our methods using both objective and subjective metrics.
For the objective evaluation, we extract speaker embeddings from generated utterances using~\cite{GE2E} and we measure Speaker Encoder Cosine Similarity (SECS), which is defined as the average cosine similarity between extracted speaker embeddings from generated utterances and target embeddings.
For the subjective evaluation, we use MUltiple Stimuli with Hidden Reference and Anchor (MUSHRA) tests with the scale 0 to 100 to perceptually assess naturalness and speaker similarity \cite{merritt2018comprehensive}.
To measure the naturalness, we ask listeners: ``Please rate the audio samples in terms of
their naturalness''. The recording from the target speaker is included among the systems to be rated, as an upper-anchor.
To measure speaker similarity, we ask listeners ``Please listen to the speaker in the reference sample first. Then rate how similar the speakers in each system sound compared to the reference speaker''. Two different recordings from the target speaker are included, one as the reference sample and the other as one of the systems to be rated, as an upper-anchor. In addition, the source speech recording is included among the systems to be rated, as a lower-anchor.

Results of the evaluations are shown in Table~\ref{tab_ZS_Flow_compare}. Differences between all systems in terms of naturalness are statistically significant with the exception of between Flow-VC and Flow-TTS with f0. This demonstrates that increasing conditioning to the flow reduces the amount of information needed to be stored in the latent space $z$. This ensures that $z$ therefore contains little information of perceptual importance, allowing us to use $z$ from the source speech or simply sample from the prior, without affecting naturalness.
This allows for performance improvements in voice conversion (i.e., retaining source speaker prosody whilst simply changing the voice identity). However, when $z$ contains information such as prosody, there is a significant improvement by using $z$ extracted from the source speech instead of sampling from the prior.
None of the differences between flow models in the speaker similarity evaluations are found to be statistically significant.
The objective evaluations find significant improvements in terms of speaker similarity by running the models in TTS mode instead of VC mode, see Table~\ref{tab:WER}. However, the differences are subtle, and this is probably the reason why comparing this to the findings from the subjective evaluations indicates that listeners are not able to perceive these improvements.
Typical TTS models do not have access to oracle f0 and vuv conditioning during inference, whereas VC models do.
Therefore, VC mode is preferred to TTS mode. 
However, \mbox{Flow-TTS} conditioned with oracle f0 and vuv is on-par with Flow-VC. 
Oracle f0 represents the upper-bound for f0 prediction. How far f0 prediction models are able to close this gap between Flow-TTS and Flow-VC is outside of the scope of this investigation.

\begin{table}[t!]
\setlength{\tabcolsep}{5pt}
\centering
\caption{Zero-shot: mean ratings from MUSHRA tests evaluating speaker similarity and naturalness of flow models.}
\vspace{-.25cm}
\begin{tabular}{ccc}
\toprule
\textbf{System} & \textbf{Speaker similarity} $\uparrow$ & \textbf{Naturalness} $\uparrow$ \\ \midrule
Source recordings  & 58.98 & -- \\
Flow-TTS with f0  & 62.53 & 68.21\\
Flow-TTS  & 62.67 & 64.54\\
Flow-VC w/o f0 & 61.96 & 65.32\\
Flow-VC  & 62.50 & 68.27 \\
Target recordings  & 75.58 & 73.38\\
\bottomrule
\end{tabular}
\label{tab_ZS_Flow_compare}
\end{table}

\subsection{Zero-shot: Comparison to the SOTA}
We evaluate our models alongside the state-of-the-art (SOTA) in VC and TTS using both objective and subjective metrics.
Results from the subjective evaluations are shown in Table~\ref{tab_ZS_Subjective_Nat_SpkSim}.
Comparing to the state-of-the-art in both naturalness and speaker similarity, Flow-VC significantly outperforms text-to-speech systems and is on par with CopyCat.
Results from the objective evaluation for speaker similarity are shown in Table~\ref{tab:WER}, and these results correlate with the findings from subjective evaluations.

Note that there is a gap between evaluated approaches and target recordings in terms of speaker similarity. This indicates that the evaluated approaches were not fully successful in achieving the zero-shot speaker identity, however Flow-VC achieves state-of-the-art results in speaker similarity to the target identity.
We hypothesise that this gap can be due to the speaker embeddings not being intuitive enough for the model to be able to fully understand everything about the target speaker identity purely from the embeddings. This is perhaps not surprising given that we need to represent the infinite possible speaker identities into a single vector. In informal analysis, we have observed that in fact different speech corpora form separate clusters in the learnt speaker embedding space. This indicates that the embeddings are also modelling factors which we are not interested in, hence further work into making these more efficient in storing purely speaker attributes is needed.

\begin{table}[t]
\centering
\caption{Zero-shot: comparison of Flow-VC with the state-of-the-art using MUSHRA tests with mean and median ratings.}
\vspace{-.2cm}
\begingroup
\setlength{\tabcolsep}{3pt}
\begin{tabular}{ccccc}%
\toprule%
    \multirow{2}{*}{\textbf{System}} &  \multicolumn{2}{c}{\textbf{Speaker similarity} $\uparrow$} & \multicolumn{2}{c}{~ ~\textbf{Naturalness} $\uparrow$ ~ } \\%
    & ~ ~ Mean ~ & Median & ~Mean~ & Median \\ \midrule%
    \multicolumn{5}{c}{\textbf{VC baselining}} \\ \midrule%
        Source recordings & ~ 56.12 & 69 & -- & --\\%
        CopyCat & ~ 59.49 & 72 & 68.63 & 76\\%
        Flow-VC & ~ 59.86 & 71 & 68.08 & 75\\%
        Target recordings & ~ 76.42 & 81 & 72.13 & 78\\ \midrule%
    \multicolumn{5}{c}{\textbf{TTS baselining}} \\ \midrule%
        Source recordings & ~ 57.27 & 71 & -- & --\\%
        Tacotron 2 & ~ 59.29 & 72 & 64.18 & 73 \\%
        EDNA-TTS & ~ 57.68 & 71 & 61.90 & 71 \\%
        Flow-VC & ~ 60.83 & 72 & 70.56 & 76\\%
        Target recordings & ~ 76.77 & 82 & 72.97 & 77\\%
        \bottomrule%
\end{tabular}%
\endgroup
\label{tab_ZS_Subjective_Nat_SpkSim}%
\end{table}

\subsection{Are we creating new voices?}
To measure the ability of creating many new voices, we need to verify that new voices are different from voices in the training set and different from each other.
We use speaker embeddings of voices in the training set and generate 120 new American speaker embeddings, see Section~\ref{sec:approach}.
We synthesize new voices with our TTS and VC models, and extract speaker embeddings from generated utterances using~\cite{GE2E}.
First, we calculate the variance of generated embeddings followed by the sum over the dimensions. Flow-VC achieves 0.28, Flow-TTS achieves 0.22 and CopyCat achieves 0.19. This indicates that flows have higher variance of generated voices, thus we focus on these.
Next, we evaluate the flow models using MUSHRA tests with a binary scale, where same speakers are represented by value 0 and different speakers by value 100.
We ask listeners ``Please listen to the speaker in the reference sample first. Then rate whether the speakers in each system sound like the same speaker as in the reference or different''.
Two different recordings from the nearest neighbour seen speaker (NN) are included, one as the reference sample and the other as one of the systems (NN) to be rated, as an anchor.
The NN is the closest speaker from the training set to the generated sample using the speaker embeddings and the cosine distance metric.
In addition, the nearest neighbour seen speaker (NN2NN) to our NN speaker is included among the systems to be rated, as an additional anchor.

Results are shown in Table~\ref{any-voice}. NN has a low score and NN2NN and flow models have significantly higher scores, which indicates that listeners are able to distinguish same and different voices, and that our new voices can indeed represent new speakers. Moreover, flow models have significantly higher scores than NN2NN, which means that new voices are further from voices in the training set than these voices in the training set from their closest voices in the training set.
This finding clearly indicates that the generated voices are indeed new and different from voices in the training data.

\begin{table}[t]
\centering
\caption{New voices: mean ratings from MUSHRA tests demonstrating that new voices are different from training voices.}
\vspace{-.2cm}
\begin{tabular}{cM{1.01cm}M{1.31cm}M{1.80cm}}
\toprule
\textbf{System} & \textbf{NN} $\downarrow$ & \textbf{NN2NN} $\uparrow$ & \textbf{New voice} $\uparrow \uparrow$ \\ \midrule
Flow-TTS & 38 & 72 & 82 \\
Flow-TTS with f0 & 31 & 69 & 78 \\
Flow-VC w/o f0 & 32 & 72 & 85 \\
Flow-VC & 25 & 75 & 81 \\
\bottomrule
\end{tabular}
\label{any-voice}%
\end{table}

\begin{figure}[t]%
\centering
\vspace{.1cm}
\includegraphics[width=0.45\textwidth]{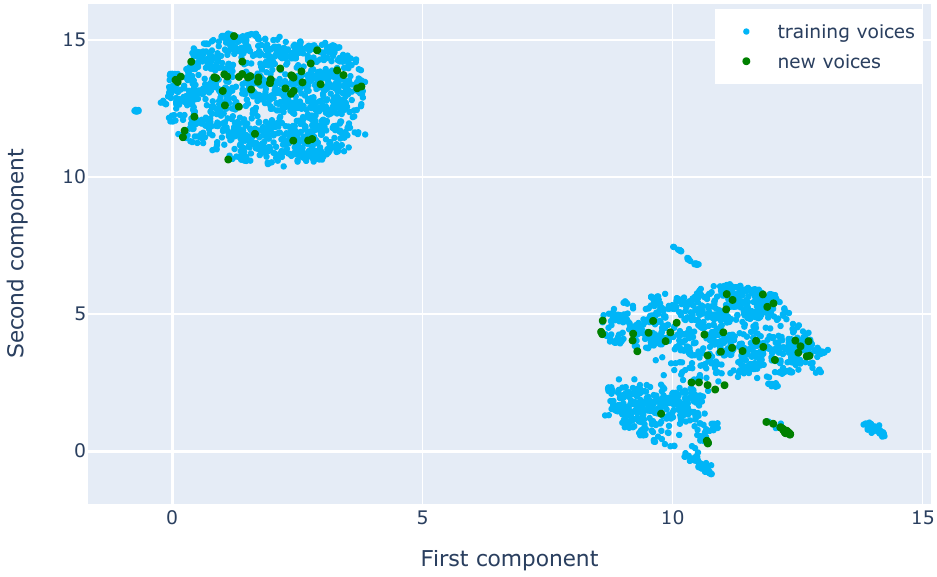}%
\vspace{-.1cm}
\caption{New voices: Visualization of speaker embedding space with both training and new voices (best viewed in color).}%
\label{fig:pca}%
\vspace{-.1cm}
\end{figure}

We also visualize the speaker embedding space with both voices in the training data and new voices.
Each speaker embedding has 256 dimensions.
We apply the Principal Component Analysis~\cite{PCA} and select 48 principal components explaining over 90\% of the total variance.
Then, we apply the Uniform Manifold Approximation and Projection~\cite{UMAP} to map them to 2D space.
Results are presented in Figure~\ref{fig:pca}.
The visualisation shows that new voices are spread across the speaker embedding space of voices from the training set, and thus this confirms that our approach can generate a variety of new voices.

\section{Conclusions}
We have presented a normalising flows-based approach and investigated its ability in TTS and VC modes to extrapolate from speakers observed during training to create unseen speaker identities.
We have comprehensively evaluated our methods and baselines in terms of intelligibility, naturalness, speaker similarity, and ability to create new voices. Extensive evaluations demonstrate that the proposed approach systematically allows to obtain state-of-the-art performance in zero-shot speech synthesis and creates various new voices, unobserved in the training set.
However none of the approaches are precise in the zero-shot scenario. We hypothesise this is due to inefficiencies in the learnt speaker embeddings. Future work on improving speaker embeddings is required to advance towards realising the ultimate zero-shot voice creation goal.

\bibliographystyle{IEEEtran}

\balance

\bibliography{main}

\begin{thebibliography}{10}
\providecommand{\url}[1]{#1}
\csname url@samestyle\endcsname
\providecommand{\newblock}{\relax}
\providecommand{\bibinfo}[2]{#2}
\providecommand{\BIBentrySTDinterwordspacing}{\spaceskip=0pt\relax}
\providecommand{\BIBentryALTinterwordstretchfactor}{4}
\providecommand{\BIBentryALTinterwordspacing}{\spaceskip=\fontdimen2\font plus
\BIBentryALTinterwordstretchfactor\fontdimen3\font minus
  \fontdimen4\font\relax}
\providecommand{\BIBforeignlanguage}[2]{{%
\expandafter\ifx\csname l@#1\endcsname\relax
\typeout{** WARNING: IEEEtran.bst: No hyphenation pattern has been}%
\typeout{** loaded for the language `#1'. Using the pattern for}%
\typeout{** the default language instead.}%
\else
\language=\csname l@#1\endcsname
\fi
#2}}
\providecommand{\BIBdecl}{\relax}
\BIBdecl

\bibitem{shen-arxiv-2020}
J.~Shen, Y.~Jia, M.~Chrzanowski, Y.~Zhang, I.~Elias, H.~Zen, and Y.~Wu,
  ``{Non-Attentive Tacotron: Robust and Controllable Neural TTS Synthesis
  Including Unsupervised Duration Modeling},'' \emph{arXiv:2010.04301}, 2020.

\bibitem{ren2021fastspeech}
Y.~Ren, C.~Hu, X.~Tan, T.~Qin, S.~Zhao, Z.~Zhao, and T.-Y. Liu, ``Fastspeech 2:
  Fast and high-quality end-to-end text to speech,'' in \emph{International
  Conference on Learning Representations, (ICLR)}, 2021.

\bibitem{ezzerg2021enhancing}
A.~Ezzerg, A.~Gabrys, B.~Putrycz, D.~Korzekwa, D.~Saez-Trigueros, D.~McHardy,
  K.~Pokora, J.~Lachowicz, J.~Lorenzo-Trueba, and V.~Klimkov, ``{Enhancing
  audio quality for expressive Neural Text-to-Speech},'' in \emph{ISCA Speech
  Synthesis Workshop (SSW)}, 2021.

\bibitem{EDP}
R.~Shah, K.~Pokora, A.~Ezzerg, V.~Klimkov, G.~Huybrechts, B.~Putrycz,
  D.~Korzekwa, and T.~Merritt, ``{Non-Autoregressive TTS with Explicit Duration
  Modelling for Low-Resource Highly Expressive Speech},'' in \emph{ISCA Speech
  Synthesis Workshop (SSW)}, 2021.

\bibitem{autovc}
K.~Qian, Y.~Zhang, S.~Chang, X.~Yang, and M.~Hasegawa{-}Johnson, ``{AutoVC:
  Zero-Shot Voice Style Transfer with Only Autoencoder Loss},'' in
  \emph{International Conference on Machine Learning (ICML)}, 2019.

\bibitem{SCGlowTTS}
E.~Casanova, C.~Shulby, E.~Gölge, N.~M. Müller, F.~S. de~Oliveira,
  A.~{Candido Jr.}, A.~da~Silva~Soares, S.~M. Aluisio, and M.~A. Ponti,
  ``{SC-GlowTTS: An Efficient Zero-Shot Multi-Speaker Text-To-Speech Model},''
  in \emph{Interspeech}, 2021.

\bibitem{speakergeneration}
D.~Stanton, M.~Shannon, S.~Mariooryad, R.~Skerry-Ryan, E.~Battenberg, T.~Bagby,
  and D.~Kao, ``Speaker generation,'' in \emph{IEEE International Conference on
  Acoustics, Speech, and Signal Processing (ICASSP)}, 2021.

\bibitem{huybrechts2021low}
G.~Huybrechts, T.~Merritt, G.~Comini, B.~Perz, R.~Shah, and J.~Lorenzo-Trueba,
  ``Low-resource expressive text-to-speech using data augmentation,'' in
  \emph{IEEE International Conference on Acoustics, Speech and Signal
  Processing (ICASSP)}, 2021.

\bibitem{Tacotron2}
J.~Shen, R.~Pang, R.~J. Weiss, M.~Schuster, N.~Jaitly, Z.~Yang, Z.~Chen,
  Y.~Zhang, Y.~Wang, R.~J. Skerry{-}Ryan, R.~A. Saurous, Y.~Agiomyrgiannakis,
  and Y.~Wu, ``Natural {TTS} synthesis by conditioning wavenet on mel
  spectrogram predictions,'' in \emph{IEEE International Conference on
  Acoustics, Speech and Signal Processing (ICASSP)}, 2018.

\bibitem{ren2019fastspeech}
Y.~Ren, Y.~Ruan, X.~Tan, T.~Qin, S.~Zhao, Z.~Zhao, and T.-Y. Liu, ``Fastspeech:
  Fast, robust and controllable text to speech,'' in \emph{Neural Information
  Processing Systems (NeurIPS)}, 2019.

\bibitem{nonparallel-2006}
A.~Mouchtaris, J.~V. der Spiegel, and P.~Mueller, ``Nonparallel training for
  voice conversion based on a parameter adaptation approach,'' \emph{IEEE
  Transactions on Audio, Speech, and Language Processing}, vol.~14, no.~3, pp.
  952--963, 2006.

\bibitem{Blow}
J.~Serr{\`a}, S.~Pascual, and C.~Segura~Perales, ``Blow: a single-scale
  hyperconditioned flow for non-parallel raw-audio voice conversion,'' in
  \emph{Neural Information Processing Systems (NeurIPS)}, 2019.

\bibitem{ours}
T.~Merritt, A.~Ezzerg, P.~Biliński, M.~Proszewska, K.~Pokora,
  R.~Barra-Chicote, and D.~Korzekwa, ``{Text-free non-parallel many-to-many
  voice conversion using normalising flows },'' in \emph{IEEE International
  Conference on Acoustics, Speech and Signal Processing (ICASSP)}, 2022.

\bibitem{li2021light}
S.~Li, B.~Ouyang, L.~Li, and Q.~Hong, ``Light-tts: Lightweight multi-speaker
  multi-lingual text-to-speech,'' in \emph{IEEE International Conference on
  Acoustics, Speech and Signal Processing (ICASSP)}, 2021.

\bibitem{NF_Intro_Review}
I.~Kobyzev, S.~J. Prince, and M.~A. Brubaker, ``Normalizing flows: An
  introduction and review of current methods,'' \emph{IEEE Transactions on
  Pattern Analysis and Machine Intelligence (TPAMI)}, vol.~43, no.~11, pp.
  3964--3979, 2021.

\bibitem{GlowTTS}
J.~Kim, S.~Kim, J.~Kong, and S.~Yoon, ``{Glow-TTS: A Generative Flow for
  Text-to-Speech via Monotonic Alignment Search},'' in \emph{Neural Information
  Processing Systems (NeurIPS)}, 2020.

\bibitem{FlowPP}
J.~Ho, X.~Chen, A.~Srinivas, Y.~Duan, and P.~Abbeel, ``Flow++: Improving
  flow-based generative models with variational dequantization and architecture
  design,'' in \emph{International Conference on Machine Learning (ICML)},
  2019.

\bibitem{FlowTTS}
C.~Miao, S.~Liang, M.~Chen, J.~Ma, S.~Wang, and J.~Xiao, ``{Flow-TTS: A
  Non-Autoregressive Network for Text to Speech Based on Flow},'' in \emph{IEEE
  International Conference on Acoustics, Speech and Signal Processing
  (ICASSP)}, 2020.

\bibitem{Flowtron}
R.~Valle, K.~J. Shih, R.~Prenger, and B.~Catanzaro, ``Flowtron: an
  autoregressive flow-based generative network for text-to-speech synthesis,''
  in \emph{International Conference on Learning Representations, (ICLR)}, 2021.

\bibitem{ParallelWaveNet}
A.~van~den Oord, Y.~Li, I.~Babuschkin, K.~Simonyan, O.~Vinyals, K.~Kavukcuoglu,
  G.~van~den Driessche, E.~Lockhart, L.~Cobo, F.~Stimberg, N.~Casagrande,
  D.~Grewe, S.~Noury, S.~Dieleman, E.~Elsen, N.~Kalchbrenner, H.~Zen,
  A.~Graves, H.~King, T.~Walters, D.~Belov, and D.~Hassabis, ``Parallel
  {W}ave{N}et: Fast high-fidelity speech synthesis,'' in \emph{International
  Conference on Machine Learning (ICML)}, 2018.

\bibitem{WaveGlow}
R.~Prenger, R.~Valle, and B.~Catanzaro, ``{WaveGlow: {A} Flow-based Generative
  Network for Speech Synthesis},'' in \emph{IEEE International Conference on
  Acoustics, Speech and Signal Processing (ICASSP)}, 2018.

\bibitem{GE2E}
L.~Wan, Q.~Wang, A.~Papir, and I.~L. Moreno, ``{Generalized End-to-End Loss for
  Speaker Verification},'' in \emph{IEEE International Conference on Acoustics,
  Speech and Signal Processing (ICASSP)}, 2018.

\bibitem{CopyCat}
S.~Karlapati, A.~Moinet, A.~Joly, V.~Klimkov, D.~Sáez-Trigueros, and
  T.~Drugman, ``{CopyCat: Many-to-Many Fine-Grained Prosody Transfer for Neural
  Text-to-Speech},'' in \emph{Interspeech}, 2020.

\bibitem{UniversalNeuralVocoder}
Y.~Jiao, A.~Gabrys, G.~Tinchev, B.~Putrycz, D.~Korzekwa, and V.~Klimkov,
  ``{Universal Neural Vocoding with Parallel WaveNet},'' in \emph{IEEE
  International Conference on Acoustics, Speech and Signal Processing
  (ICASSP)}, 2021.

\bibitem{VCTK_OLD}
C.~Veaux, J.~Yamagishi, and K.~MacDonald, ``{CSTR VCTK Corpus: English
  Multi-speaker Corpus for CSTR Voice Cloning Toolkit},'' {The Centre for
  Speech Technology Research (CSTR), University of Edinburgh}.

\bibitem{VCTK}
C.~Veaux, J.~Yamagishi, and S.~King, ``{The voice bank corpus: Design,
  collection and data analysis of a large regional accent speech database},''
  in \emph{International Conference Oriental COCOSDA held jointly with 2013
  Conference on Asian Spoken Language Research and Evaluation
  (O-COCOSDA/CASLRE)}, 2013.

\bibitem{merritt2018comprehensive}
T.~Merritt, B.~Putrycz, A.~Nadolski, T.~Ye, D.~Korzekwa, W.~Dolecki,
  T.~Drugman, V.~Klimkov, A.~Moinet, A.~Breen, R.~Kuklinski, N.~Strom, and
  R.~Barra-Chicote, ``Comprehensive evaluation of statistical speech waveform
  synthesis,'' in \emph{IEEE Spoken Language Technology Workshop (SLT)}, 2018.

\bibitem{PCA}
K.~F. Pearson, ``{LIII. On lines and planes of closest fit to systems of points
  in space},'' \emph{The London, Edinburgh, and Dublin Philosophical Magazine
  and Journal of Science}, vol.~2, no.~11, pp. 559--572, 1901.

\bibitem{UMAP}
L.~McInnes, J.~Healy, N.~Saul, and L.~Großberger, ``{UMAP: Uniform Manifold
  Approximation and Projection},'' \emph{Journal of Open Source Software},
  vol.~3, no.~29, p. 861, 2018.

\end{thebibliography}

\end{document}